\documentclass[lettersize,journal]{IEEEtran}
\usepackage{graphicx}
\usepackage{amssymb}
\usepackage{amsmath}
\usepackage{mathtools}
\usepackage{cite}
\usepackage{stfloats}
\usepackage{epstopdf}
\usepackage{psfrag}
\usepackage[mathscr]{euscript}
\usepackage{acronym}  % make an acronym
\usepackage{booktabs}
\usepackage[table]{xcolor}
\usepackage{tikz}
\usepackage{url}
\usepackage[many]{tcolorbox}
\usepackage{algorithm}
\usepackage[noend]{algpseudocode}
\usepackage{kotex}
\usepackage{float}
\def\BibTeX{{\rm B\kern-.05em{\sc i\kern-.025em b}\kern-.08em
		T\kern-.1667em\lower.7ex\hbox{E}\kern-.125emX}}
\usepackage[breaklinks, plainpages=false, pdfpagelabels=false, bookmarksnumbered=false]{hyperref} % create hyperlinks

\usepackage{breakurl}

\acrodef{BS}{base station}
\acrodef{EC}{edge computing}
\acrodef{RL}{reinforcement learning}
\acrodef{UE}{user equipment}
\acrodef{LoS}{line-of-sight}
\acrodef{NLoS}{non-line-of-sight}
\acrodef{UDN}{ultra-dense network}

\acrodef{PPP}{Poisson point process}
\acrodefplural{PPP}[PPPs]{Poisson point processes}

\acrodef{PDF}{probability density function}
\acrodef{CCDF}{complementary cumulative distribution function}
\acrodef{SIR}{signal-to-interference ratio}
\acrodef{SINR}{signal-to-interference-plus-noise ratio}
\acrodef{PGFL}{probability generating functional}
\acrodef{ASE}{area spectral efficiency}

\acrodef{QoS}{quality of service}
\acrodef{HetNet}{heterogeneous network}

\acrodef{CSI}{channel state information}
\acrodef{RF}{radio frequency}
\acrodef{QoS}{quality-of-service}

\acrodef{C-RAN}{Cloud-Radio Access Network}

\acrodef{3GPP}{third generation partnership project}
\acrodef{CoMP}{coordinated multi-point}
\acrodef{JT}{joint transmission}
\acrodef{5G}{fifth-generation}
\acrodef{CoMPJT}{\ac{CoMP} based joint transmission}
\acrodef{DPS}{dynamic transmission point selection}

\usepackage{color}
\usepackage{dsfont}
\usepackage{bbm}

\newcommand{\Ri}[1]{r_{#1}}

\newcommand{\AngToDist}[1]{\frac{180}{\pi}\text{arctan}\left(\frac{\ell}{\Ri{i}}\right)}

\renewcommand\IEEEkeywordsname{Keywords}

\newcommand{\red}[1]{{\textcolor[rgb]{0,0,0}{#1}}}
\newcommand{\blue}[1]{{\textcolor[rgb]{0,0,0}{#1}}}

\acrodef{AI}{artificial intelligence}
\acrodef{MBS}{macro base station}
\acrodef{SBS}{small base station}
\acrodef{OFDMA}{orthogonal frequency-division multiple access}
\acrodef{LoS}{line-of-sight}
\acrodef{NLoS}{non-line-of-sight}
\acrodef{AWGN}{additive white Gaussian noise}
\acrodef{SINR}{signal-to-interference-plus-noise ratio }
\acrodef{PPO}{proximal policy optimization}
\acrodef{MARL}{multi-agent reinforcement learning}
\acrodef{DDPG}{deep deterministic policy gradient}
\acrodef{MDP}{markov decision process}
\acrodef{DRL}{deep reinforcement learning}
\acrodef{LLM}{large language model}
\acrodef{SLM}{small language model}
\acrodef{O-RAN}{open radio access network}
\acrodef{RAN}{radio access network}
\acrodef{RIC}{\ac{RAN} intelligent controller}
\acrodef{RU}{radio unit}
\acrodef{CU}{central unit}
\acrodef{DU}{distributed unit}
\acrodef{near-RT}{near-real-time}
\acrodef{non-RT}{non-real-time}
\acrodef{near-RT RIC}{near-real-time \ac{RIC}}
\acrodef{non-RT RIC}{non-real-time \ac{RIC}}
\acrodef{CNN}{convolutional neural network}
\acrodef{IAB}{integrated access and backhaul}
\acrodef{gNB}{next generation node B}
\acrodef{RL}{reinforcement learning}
\acrodef{ML}{machine learning}
\acrodef{RAG}{retrieval-augmented generation}
\acrodef{PRB}{physical resource block}
\acrodef{SMO}{service management and orchestration}
\acrodef{PG Loss}{policy gradient loss}
\acrodef{KL}{Kullback–Leibler}
\acrodef{MSE}{mean squared error}
\acrodef{SFT}{supervised fine-tuning}
\acrodef{RLFT}{reinforcement learning-based fine-tuning}
\acrodef{EPA}{equal power allocation}
\acrodef{SER}{successful extraction rate}
\acrodef{SMO}{service management and orchestration}
\acrodef{SCA}{Successive Convex Approximation}

\begin{document}

\newcommand{\paperTitle}{
EvoRIC: Reinforcement Learning Fine-Tuned LLM-empowered RAN Intelligent Control 
\\Toward Autonomous O-RAN 
}
%Large Language Model Empowered Hierarchical Intelligent Control for O-RAN
%Hierarchical RAN Intelligent Controller Leveraging Large Language Models in O-RAN
%LLM-Powered Hierarchical Controller for Dynamic Resource Management in O-RAN

%---------------------------------------------------------------------------%
%                     title, title footnote, header                         %
%---------------------------------------------------------------------------%

%\twocolumn

% paper title
\title{\paperTitle}
% \title{Distributed Secrecy in \\Multilevel Wireless Networks}

\author{Lingyan Bao,~\IEEEmembership{Graduate Student Member,~IEEE}, Jemin Lee,~\IEEEmembership{Senior Member,~IEEE},\\and Tony Q.S. Quek, ~\IEEEmembership{Fellow,~IEEE}
\thanks{
	Manuscript submitted 23 April 2026; revised 7 August 2026.
}
\thanks{
	Corresponding author is J. Lee.
}
\thanks{
	J. Lee and L. Bao are with the School of Electrical and Electronic Engineering, Yonsei University, Seoul 03722, South Korea (e-mail: jemin.lee@yonsei.ac.kr; lingyan@yonsei.ac.kr). 
	
	T. Q. S. Quek is with the Information System Technology and Design, Singapore University of Technology and Design, Singapore 487372 (e-mail:tonyquek@sutd.edu.sg). 
}
%\thanks{
%			J. Lee is with the School of Electrical and Electronic Engineering, Yonsei
%			University, Seoul 03722, South Korea (e-mail: jemin.lee@yonsei.ac.kr).
%		}
}
%\thanks{
%}
%
%\thanks{
%		J. Lee is with the School of Electrical and Electronic Engineering, Yonsei
%		University, Seoul 03722, South Korea (e-mail: jemin.lee@yonsei.ac.kr).
%		%J. Lee is with the School of Electrical and Electronic Engineering, Yonsei
%		%University, Seoul 03722, South Korea (e-mail: jemin.lee@yonsei.ac.kr).
%	}
%
%}

\maketitle %% make the title area

\setcounter{page}{1}
\renewcommand\IEEEkeywordsname{Index Terms}

%%---------------------------------------------------------------------------%
%%                           abstract and key words                          %
%%---------------------------------------------------------------------------%
\begin{abstract}
Despite recent advances in applying \ac{AI} techniques to \ac{RAN}, critical challenges remain: traditional \ac{ML} algorithms suffer from limited generalization across varying network topologies, whereas general-purpose \acp{LLM} face high computational demands and lack domain-specific knowledge. To address these gaps, this article introduces the evolving \ac{RIC} (EvoRIC) framework, a hierarchical architecture that enables continuous evolution by leveraging a \ac{non-RT RIC} for global model updates and a \ac{near-RT RIC} for local execution, dynamically empowering \acp{LLM} with domain-specific decision-making capabilities. Within this framework, we employ a \ac{RLFT} mechanism where \blue{an \ac{LLM}} operates as an actor within a \ac{PPO} agent. By leveraging the interaction \blue{tuples} collected from the wireless environment, the \blue{\ac{LLM}'s} parameters are iteratively updated to align semantic reasoning with rigorous network performance objectives. We evaluate the generalization and efficacy of the proposed EvoRIC framework within \ac{IAB} networks, and finally, discuss the open challenges and future directions of the EvoRIC framework toward realizing autonomous O-RAN.
\end{abstract}


\acresetall
\section{Introduction}
The rapidly advancing landscape of 6G has positioned \ac{AI} algorithms as a foundational element for driving innovations in advanced \ac{RAN}. Within this context, the concept of \ac{AI}-\ac{RAN}, leveraging \ac{AI} to enhance \ac{RAN} performance, has gained significant traction. Specifically, the \ac{AI}-for-\ac{RAN} paradigm focuses on integrating intelligence directly into the network infrastructure to optimize critical metrics such as spectral efficiency and capacity. 
A key enabler of this vision is the \ac{O-RAN} architecture, which introduces the \ac{RIC} to decouple control functionalities from hardware. This architectural shift paves the way for autonomous O-RAN, a paradigm aiming for zero-touch, self-optimizing network operations. In \ac{O-RAN}, \ac{AI} algorithms are deployed as microservices known as xApps (in \ac{near-RT RIC}) and rApps (in \ac{non-RT RIC}), enabling programmable network optimization across different control timescales \cite{polese2023understanding}.

Among various \ac{AI} approaches, \ac{RL} has emerged as a dominant tool for solving complex resource allocation problems in wireless communications. To identify near-optimal policies, an \ac{RL} agent interacts with the environment, optimizing network parameters to maximize long-term rewards. \blue{In \cite{qiao2025resource}, the \ac{RL} agent is trained in the \ac{non-RT RIC} and subsequently deployed in the \ac{near-RT RIC} to perform resource management for network slicing.} However, traditional \ac{RL} algorithms struggle to adapt efficiently to the continuously varying action and state spaces inherent in dynamic wireless environments. This limitation often necessitates frequent retraining or the maintenance of multiple model versions for different network topologies, posing a significant challenge to scalable deployment in heterogeneous networks.

Recently, \acp{LLM} have demonstrated remarkable capabilities in natural language understanding, reasoning, and complex decision-making, driven by their extensive training on massive datasets. 
The wireless community has begun to explore the potential of \acp{LLM} for network optimization tasks, such as resource allocation \cite{lee2026llm,zhou2025prompting}. 
Early attempts utilized zero-shot learning or in-context learning, where the \ac{LLM} infers policies based on provided examples (prompts) without parameter updates. 
\blue{Notably, hierarchical \ac{RIC} architectures, i.e., the LLM-$h$RIC framework \cite{bao2026llm}, deploy an \ac{LLM} in the \ac{non-RT RIC} to provide high-level policy guidance based on global network information, while a \ac{DDPG} agent in the \ac{near-RT RIC} performs the resource optimization.} 
% However, this framework relies strictly on zero-shot prompting with general-purpose models and does not involve domain-specific fine-tuning, leaving the model's reasoning constrained by its pre-trained knowledge.}
More advanced approaches employ \ac{SFT} to adapt \acp{LLM} for specific tasks, ranging from channel estimation and reconstruction \cite{11071329} to technical problems summarization \cite{lin2025empowering}. 
Furthermore, \ac{LLM}-based agents have been applied to packet analysis \cite{kan2024mobile} and \ac{O-RAN} knowledge retrieval \cite{gajjar2025oran}.

Despite these advancements, the application of \ac{LLM}-based techniques in wireless networks faces several critical challenges:
1) in-context learning struggles to guarantee consistent performance as providing high-quality and context-aware services in rapidly changing wireless channels is inherently difficult;
2) the widespread adoption of \ac{SFT} remains constrained by the scarcity of comprehensive and expert-labeled decision-making datasets for most wireless scenarios;
3) although general-purpose \acp{LLM} possess superior generalization capabilities compared to traditional \ac{AI}, their high computational complexity hinders near-RT decision-making; and
4) without rigorous alignment, \acp{LLM} are prone to generating syntactically correct but physically unexecutable control actions.

To overcome these critical challenges, recent research has started investigating the synergy between \acp{LLM} and \ac{RL}, specifically formulating \acp{LLM} as adaptive agents capable of self-optimization via environmental feedback \cite{schmied2025llms}. 
This convergence offers a unique opportunity to combine the generalization capabilities of foundation models with the autonomous optimization power of \ac{RL}, thereby circumventing the need for expensive labeled datasets.

Motivated by these opportunities and challenges, we propose a novel framework for \ac{O-RAN}, termed evolving RIC (EvoRIC), which improves its control policy through closed-loop interaction with the wireless environments. 
In this framework, a compact \ac{LLM} (or \ac{SLM}) operates as the \ac{RL} actor within the \ac{non-RT RIC}, where it is fine-tuned using a critic-guided policy gradient approach. 
Subsequently, the optimized models are deployed to the \ac{near-RT RIC} to execute near-RT resource allocation in the wireless environment. 
\blue{The motivation for introducing an LLM-based control arises from supporting the structural variability of control problems. Across heterogeneous deployments, number of controlled network entities, length of the network-state description, dimension of the control actions, as well as associated operational constraints, the network can have various. }
To facilitate continuous learning, interaction trajectories are aggregated from the \ac{near-RT RIC} and fed back to the \ac{non-RT RIC} via standard open interfaces (e.g., O1).
The key contributions of this work are summarized as follows:

\begin{itemize}
    \item The proposed EvoRIC framework leverages the inherent reasoning and transfer capabilities of \acp{LLM}. This facilitates generalization across diverse network scenarios (e.g., varying numbers of BSs or users) without the need for architecture reconstruction.
    
    \item We adopt an \ac{RL}-driven fine-tuning mechanism that enables the \ac{LLM} to optimize its policy directly through environmental interaction. This approach circumvents the data bottleneck associated with the scarcity of expert-labeled datasets in traditional supervised methods.
    
    \item We evaluate the feasibility and performance of EvoRIC within \ac{IAB} networks to demonstrate the superiority of our fine-tuned compact model over massive general-purpose models.
\end{itemize}

\begin{figure*}
	\centering
	\includegraphics[width=0.9\linewidth]{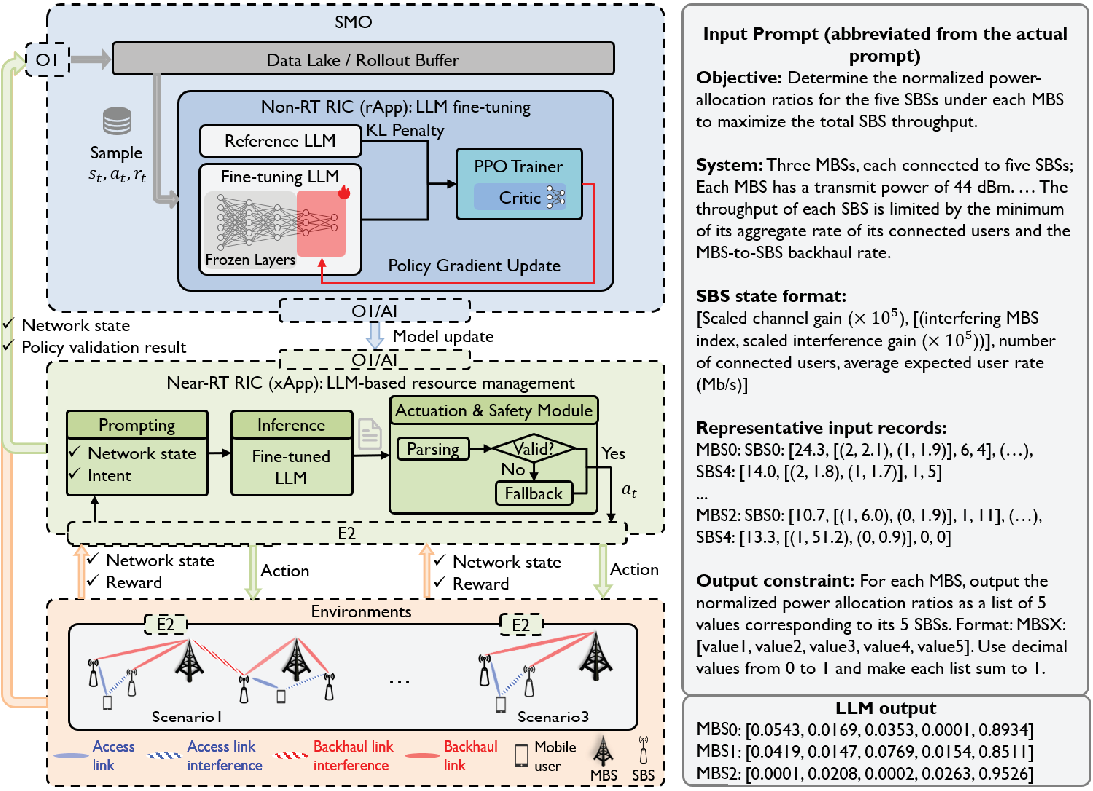}
	\caption{Proposed EvoRIC framework: a closed-loop architecture integrating a \ac{non-RT RIC} for global model fine-tuning and a \ac{near-RT RIC} for near-real-time inference, together with an abbreviated prompt template and representative \ac{LLM} output for power allocation.}
	\label{fig:RLFT-LRIC}
\end{figure*}

\section{Proposed EvoRIC Framework}
\blue{The \ac{O-RAN} separates \ac{RAN} control into the \ac{non-RT RIC} and \ac{near-RT RIC}, which operate with different responsibilities and timescales \cite{polese2023understanding}. The \textit{\ac{non-RT RIC}} is
integrated with the \ac{SMO}
framework and hosts rApps for long-term policy optimization and \ac{AI} model lifecycle management, in a timescale of greater than $1$ second. In contrast, the \textit{\ac{near-RT RIC}} hosts xApps that process \ac{near-RT} measurements 
collected from \ac{RAN} nodes and generate control actions for
resource management in a timescale of $10$~ms to $1$~s. These functions interact through \textit{open interfaces} such as E2 for measurements and control messages exchange between the \ac{near-RT RIC} and \ac{RAN} nodes, A1 for policies and model information exchange between the \ac{non-RT RIC} and \ac{near-RT RIC}, and O1 for management and data exchange between the \ac{SMO} and \ac{RAN} nodes.} 

{In this section, we describe the} proposed EvoRIC framework for decision making in \ac{O-RAN}, as illustrated in Fig. \ref{fig:RLFT-LRIC}. 
Architecturally, the framework decouples the control loop into two distinct and yet interconnected domains:
\begin{itemize}
    \item[1)] \textit{Non-RT RIC} for \ac{LLM} fine-tuning: functioning as a training domain that orchestrates the resource-intensive \ac{LLM} fine-tuning (e.g., the \ac{PPO} trainer) using aggregated historical data.
    \item[2)] \textit{Near-RT RIC} for \ac{LLM} inference: serving as an inference domain, responsible for executing near-RT decision-making at a network edge.
    % responsible for executing near-RT decision-making at a network edge.
\end{itemize}
By aggregating telemetry from distributed \acp{near-RT RIC}, the \ac{non-RT RIC} establishes a closed data loop between centralized fine-tuning and edge execution. \blue{At each \ac{near-RT RIC}, the deployed \ac{LLM} processes a prompt containing the intent, network state, and constraints to generate a corresponding control policy. The resulting \blue{interaction tuple}, consisting of the input prompt, generated policy, and environmental feedback, is returned to the \ac{non-RT RIC} for subsequent fine-tuning. Learning from \blue{interaction tuples} collected under diverse network conditions helps the \ac{LLM} adapt to heterogeneous network scenarios. The updated \ac{LLM} is then redeployed to the \acp{near-RT RIC}. Repeating this collect-update-deploy cycle enables the \ac{LLM} policy to evolve from new network experiences, without requiring manually labeled optimal actions. This feedback-driven cycle is the core mechanism underlying the continuous evolution of EvoRIC. Detailed descriptions of each component are provided in the following subsections.}

\subsection{Non-RT RIC for \Ac{LLM} Fine-Tuning} 
\blue{The \ac{non-RT RIC} leverages its global visibility to fine-tune the \ac{LLM} using the data collected from \acp{near-RT RIC}. The \ac{RL}-based fine-tuning workflow proceeds as follows:}
\begin{itemize}
    \item[1)] \red{\emph{Data Aggregation and Rollout Collection:} The \ac{non-RT RIC} aggregates interaction logs from distributed \acp{near-RT RIC} via the O1 interface. Each interaction record is structured as a transition tuple $(s_t,a_t,r_t)$ and stored in a rollout buffer $\mathcal{D}$, where $s_t$ denotes the input prompt, $a_t$ is the policy generated by the \ac{LLM}, and $r_t$ is the reward returned by the environment.}
    \item[2)] \emph{RL-based \ac{LLM} Fine-Tuning:} To align the \ac{LLM}'s semantic reasoning with network performance objectives, the \ac{non-RT RIC} executes a \ac{RL} training loop \cite{schmied2025llms}. The detailed mechanism includes three key aspects:
    \par \textbf{\textit{a. Shift from \ac{SFT} to Autonomous Exploration:}} A critical innovation of our approach is the shift from \textit{supervised instruction following} to \textit{autonomous exploration}. \blue{Unlike \ac{SFT}, which relies on scarce and costly expert-labeled data, the \ac{RL}-driven mechanism enables the \ac{LLM} to explore candidate actions, receive rewards based on their network performance, and update its policy toward higher-reward decisions.}
    \par \textbf{\textit{b. Composite Reward Function:}} 
    To ensure performance and executability,
    %To ensure the generated policies are both performant and executable, 
    we design a composite reward function that simultaneously optimizes objective adherence (e.g., throughput) and format compliance. This mechanism significantly reduces the hallucination rate (i.e., non-executable outputs), ensuring that even a fine-tuned compact model can generate reliable control actions.
    \par \textbf{\textit{c. Dual-Model Mechanism for Stability:}} To improve training stability and prevent the model from deviating excessively from its pre-trained knowledge (i.e., catastrophic forgetting), a dual-model mechanism is employed. For each sampled prompt, the policy \ac{LLM} generates an action sequence, while the reference model evaluates the same sequence and provides the token-level reference log-probabilities. \blue{The resulting \ac{KL}-divergence penalty is incorporated into the PPO reward as a soft trust-region regularizer to discourage large policy deviations that may cause unstable updates\cite{liu2025rethinking}.}
    % It therefore balances task-specific optimization with cross-scenario generalization and transferability ~\cite{liu2025rethinking}.}
    \item[3)] \emph{Model Update and Deployment:} Periodically or upon the convergence of the training epoch, the updated model is deployed to \acp{near-RT RIC} via the A1 interface. Capitalizing on the generalization capabilities of the \ac{LLM}, a common model fine-tuned in the \ac{non-RT RIC} can be deployed across multiple \acp{near-RT RIC} serving diverse network topologies. 
\end{itemize}

\subsection{Near-RT RIC for \Ac{LLM} Inference} \blue{The \ac{near-RT RIC} operates as a distributed inference engine, executing the fine-tuned \ac{LLM} for \ac{near-RT} radio resource management.} By acting as a tactical execution layer, it utilizes the \ac{LLM}'s generalized reasoning capabilities to adaptively optimize network parameters in response to fluctuating traffic patterns and topological changes. The inference workflow proceeds as follows: 
\begin{itemize}
    \item[1)] \emph{Intent-and-Policy Driven Prompting:} \red{The \ac{near-RT RIC} aggregates network states from the \ac{gNB} via the E2 interface. Depending on the optimization task, these states may include, for example, channel conditions, buffer states, and resource availability.} To bridge the modality gap between continuous signals and discrete tokens, these numerical metrics are serialized into a structured text and combined with the optimization objective. The resulting prompt further includes two control directives: 
    1) a \textit{high-level service intents} (e.g., `maximizing network throughput'), which defines the optimization direction; and 
    2) \textit{explicit policy specifications} (e.g., `output a list of $N$ power levels'), which strictly define the required output format. This composite prompt guides the \ac{LLM} to generate policies that are not only optimal for the intent but also syntactically valid and structurally compliant with the network requirements.
    \item[2)] \emph{Near-RT Inference:} The fine-tuned \ac{LLM} processes the prompt to generate a structured textual response that encapsulates the intended policy action. By utilizing a model with small size, we can make this inference process complete within a sub-second control window, satisfying the strict latency constraints of the \ac{near-RT RIC}. 
    \item[3)] \emph{Actuation \& Safety Verification Module:} A deterministic parsing mechanism processes the generated textual response to extract the policy actions (e.g., power levels, beam configurations) from the textual response. Before execution, the extracted policy actions undergo a safety verification check to ensure compliance with physical constraints (e.g., the maximum power budget). If the generated policy action fails this validation, a default fallback strategy (e.g., a conservative baseline configuration) is triggered to maintain network stability. Only validated policy actions are encapsulated into E2 control messages and transmitted to the \ac{gNB} for execution.
    \blue{\item[4)] \emph{Feedback Collection:} After execution, the input prompt, the generated action, and the corresponding environment or validation feedback are returned from \ac{near-RT RIC} to the \ac{non-RT RIC} through O1 and stored in the rollout buffer for subsequent fine-tuning. This feedback completes the collect-update-deploy loop between \ac{near-RT} execution and \ac{non-RT} model evolution.}
\end{itemize}

\section{EvoRIC Framework for Power Allocation} In this section, we present a use case based on \ac{IAB} networks to verify the effectiveness of the proposed EvoRIC framework. This scenario was selected as a foundational proof-of-concept, as resource allocation in multi-cell \ac{IAB} networks is a recognized challenge requiring robust adaptability. A critical hurdle in \ac{IAB} networks is the diverse deployment configurations of \ac{MBS} and \ac{SBS} nodes. This variability renders traditional machine learning approaches ineffective, as they typically use the rigid models, trained separately for each specific network configuration. This problem effectively demonstrates the core strength of our framework: it leverages the inherent generalization capabilities of the \ac{LLM} to adapt to varying network structures. Furthermore, the \ac{RL} trainer automatically guides the \ac{LLM} towards optimal decision-making, ensuring that the model's generative capacity is strictly aligned with the rigorous performance objectives of the wireless network.

\subsection{System Model}\label{subsec:3}
As illustrated in Fig. \ref{fig:RLFT-LRIC}, the \ac{IAB} network consists of multiple \acp{MBS}, distributed in the network. Each \ac{MBS} connects to a cluster of $N$ \acp{SBS} via wireless backhaul. A set of mobile users served by these \acp{SBS} follow a Gaussian-Markov mobility model. 
The \blue{\acp{SBS}} allocate a fixed power to its connected users. However, interference exists among \acp{SBS} connected to different \ac{MBS} that use the same sub-carrier. 
The channel model considers both \ac{LoS} and \ac{NLoS} path loss characteristics, where the \ac{LoS} probability decays exponentially with the distance, alongside independent Nakagami-m fading small-scale fading.
% The probability of a \ac{LoS} connection is modeled as $P^{\text{Los}} = \exp(-d/\rho)$, where $d$ represents the distance between \ac{MBS} and \ac{SBS}, and $\rho$ is a \ac{LoS} range constant. Furthermore, the small-scale fading is modeled as independent Nakagami-m fading.

In this use case, the total channel bandwidth $W$ is assigned to the backhaul link $W\alpha$ and the access links $W(1-\alpha)$ based on a proportion factor $\alpha \in [0,1]$. 
Consequently, the effective data rate for each user is constrained by the bottleneck formed between the backhaul capacity and the aggregate access demand. The system objective is to maximize the total network throughput, defined as the summation of these bottleneck-limited rates across all nodes. This is achieved by optimizing the power allocation from each \ac{MBS} to its associated \acp{SBS}, strictly subject to the maximum power budget constraint of each individual \ac{MBS}.

\subsection{EvoRIC for power allocation} 
In this section, the proposed EvoRIC framework is deployed to solve the power allocation problem in \ac{IAB} networks. \red{The \acp{MBS} are associated with a \ac{near-RT RIC} for \ac{near-RT} inference and data collection, while a centralized \ac{non-RT RIC} aggregates global information for \ac{RL}-based fine-tuning.}

\subsubsection{MDP Formulation}
To facilitate the \ac{LLM} fine-tuning via \ac{RL}, we model the power allocation problem as a Markov Decision Process (MDP) and define its state, action, and reward components, denoted by $(\mathcal{S}, \mathcal{A}, \mathcal{R})$, as follows:
\begin{itemize}
\item State Space ($\mathcal{S}$): \red{The state $s_t$ denotes the prompt provided to the LLM at control interval $t$. As illustrated in Fig.~\ref{fig:RLFT-LRIC}, it comprises three components: 1) the optimization objective, i.e., maximizing total throughput; 2) the network context, including the numbers of \acp{MBS} and \acp{SBS}, average channel gains, user information, and interference sources; and 3) the action specification, which requires $N$ normalized power-allocation values. The network measurements are inserted into this template to construct $s_t$.}
\item Action Space ($\mathcal{A}$): The action $a_t$ corresponds to the power allocation vector $p_{m}=(p_{m,1},\cdot\cdot\cdot,p_{m,N})$, where $p_{m,n} \in [0,1]$ represents the normalized power ratio allocated to the $n$-th \ac{SBS}.
\item Reward Function ($\mathcal{R}$): The reward $r_t = r_{\text{th}} + r_{\text{format}}$ is designed to maximize throughput while enforcing strict format compliance. It consists of the normalized environment reward (throughput) and a penalty term. To ensure system operability, a substantial penalty (e.g., $r_{\text{format}} = -5$) is imposed if the \ac{LLM} fails to generate a valid, parsable numerical list.
\end{itemize}

\subsubsection{\ac{non-RT RIC} for \ac{LLM} Fine-tuning}
\red{
The \ac{non-RT RIC} iteratively fine-tunes the \ac{LLM} using \ac{PPO}.
The \ac{PPO} trainer computes the policy gradient loss from the interaction tuples $(s_t, a_t, r_t)$ sampled from the rollout buffer $\mathcal{D}$ and updates the trainable \ac{LLM} parameters to maximize the expected reward, while a KL-divergence penalty relative to the frozen reference model limits policy drift and improves stability.
}

\subsubsection{Near-RT RIC for \ac{LLM} Inference}
In the inference domain, the fine-tuned \ac{LLM} is deployed within the \ac{near-RT RIC} to execute power allocation.
To bridge the gap between numerical network states and the language model, a structured prompt template is utilized. \red{A representative prompt example, together with the corresponding power-allocation output, is presented on the right-hand side of Fig.~\ref{fig:RLFT-LRIC}.}

\subsubsection{\textit{Actuation \& Safety Verification Module}} To deploy \ac{LLM} safely within critical infrastructure, we implement a hybrid actuation mechanism that strictly isolates reasoning from execution. 
The workflow proceeds in three stages as follows:
\begin{itemize}
    \item \emph{Extraction:} A deterministic parsing algorithm (e.g., regex-based extraction) processes the \ac{LLM}'s textual response to isolate the numerical power allocation vector by discarding any extraneous linguistic tokens.
    \item \emph{Verification:} This extracted vector undergoes a safety verification check (``guardrail'') to validate compliance with physical constraints, specifically checking if the total power budget satisfies $\sum p_{n} \le P_{max}$ and if the output dimensionality matches the network topology.
    \item \emph{Fail-Safe Execution:} If the validation fails due to model hallucination or format errors, the system triggers an \emph{fail-safe protocol} that reverts to an \ac{EPA} strategy (i.e., the conservative baseline).
\end{itemize}
Consequently, this mechanism guarantees that only physically valid and safe control actions, whether generated by the \ac{LLM} or the fallback logic, are encapsulated into E2 control messages and transmitted to the \ac{gNB}, ensuring system availability and stability. \blue{After execution, the interaction tuple $(s_t,a_t,r_t)$ is stored in the rollout buffer for subsequent fine-tuning.}

\subsection{Experimental Results}
Our simulations are conducted on a high-performance workstation equipped with a single NVIDIA RTX 4090 GPU. \blue{We utilize the Llama-3.2-3B-instruct as the backbone model of EvoRIC \cite{grattafiori2024llama}. }
During the fine-tuning process, we implement a linear temperature decay schedule annealing from $1.5$ to $0.6$ combined with top\_p of 0.95. 
The \ac{PPO} algorithm is driven by the Adam optimizer with a learning rate of $10^{-5}$ and a batch size of $16$. 
\red{The model is loaded in bfloat16 precision to reduce memory usage. We evaluate the proposed EvoRIC with different fine-tuning depths by training the final $L$ transformer blocks, denoted as EvoRIC-L1 ($L=1$), EvoRIC-L2 ($L=2$), and EvoRIC-L3 ($L=3$). Compared with EvoRIC-L1, EvoRIC-L2 improves throughput by $11.52\%$, while the improvement from $L=2$ to $L=3$ is marginal (e.g., $1.20\%$).
%EvoRIC-L3 provides only a further $1.20\%$ gain. 
Therefore, EvoRIC-L2 is adopted in all subsequent simulations to balance training overhead and network performance.}

\subsubsection{Baselines} 
\blue{We compare our fine-tuned compact 3B model with massive cloud-based models (e.g., DeepSeek and Gemini).} It is important to clarify that this comparison highlights the trade-off between \textit{general-purpose reasoning} and \textit{domain-specific efficiency}. While massive models possess vast knowledge, their computational requirements make deployment at the \ac{near-RT RIC} impractical. Our goal is to demonstrate that a specialized compact model, feasible for edge deployment, can surpass the performance of cloud-based giants in specific vertical tasks, effectively validating the `Small Model with Fine-Tuning' paradigm for O-RAN. The baselines are described as follows:

\begin{itemize}
    \item \blue{Llama-3.2-3B-Instruct without fine-tuning \blue{(Llama3B-NoFT)}}: The pre-trained model without any domain-specific fine-tuning. This baseline serves to quantify the gain achieved by our \ac{RL}-based fine-tuning mechanism.
    
    \item General-Purpose Large Models (DeepSeek \& Gemini): \blue{We evaluate two massive, cloud-based models: DeepSeek-V3 (671B parameters) \cite{liu2024deepseek} and Google Gemini 1.5 Flash \cite{team2024gemini}, accessed via APIs.} To ensure a rigorous comparison, we applied comprehensive prompt engineering (incorporating task context and strict format constraints) to maximize their zero-shot capabilities. This baseline represents the representative general-purpose \ac{LLM} baselines without parameter updates.
    
    \item \ac{EPA}: A conventional heuristic scheme where power is equally distributed among all \acp{SBS}. This serves as a standard reference for traditional engineering approaches.

    \item \blue{\ac{SCA}-based Algorithm: A model-based numerical optimization method that iteratively approximates the original non-convex power-allocation problem using tractable convex subproblems implemented in CVXPY \cite{diamond2016cvxpy}.}
\end{itemize}

\begin{figure}[t]
	\centering	\includegraphics[width=0.9\linewidth]{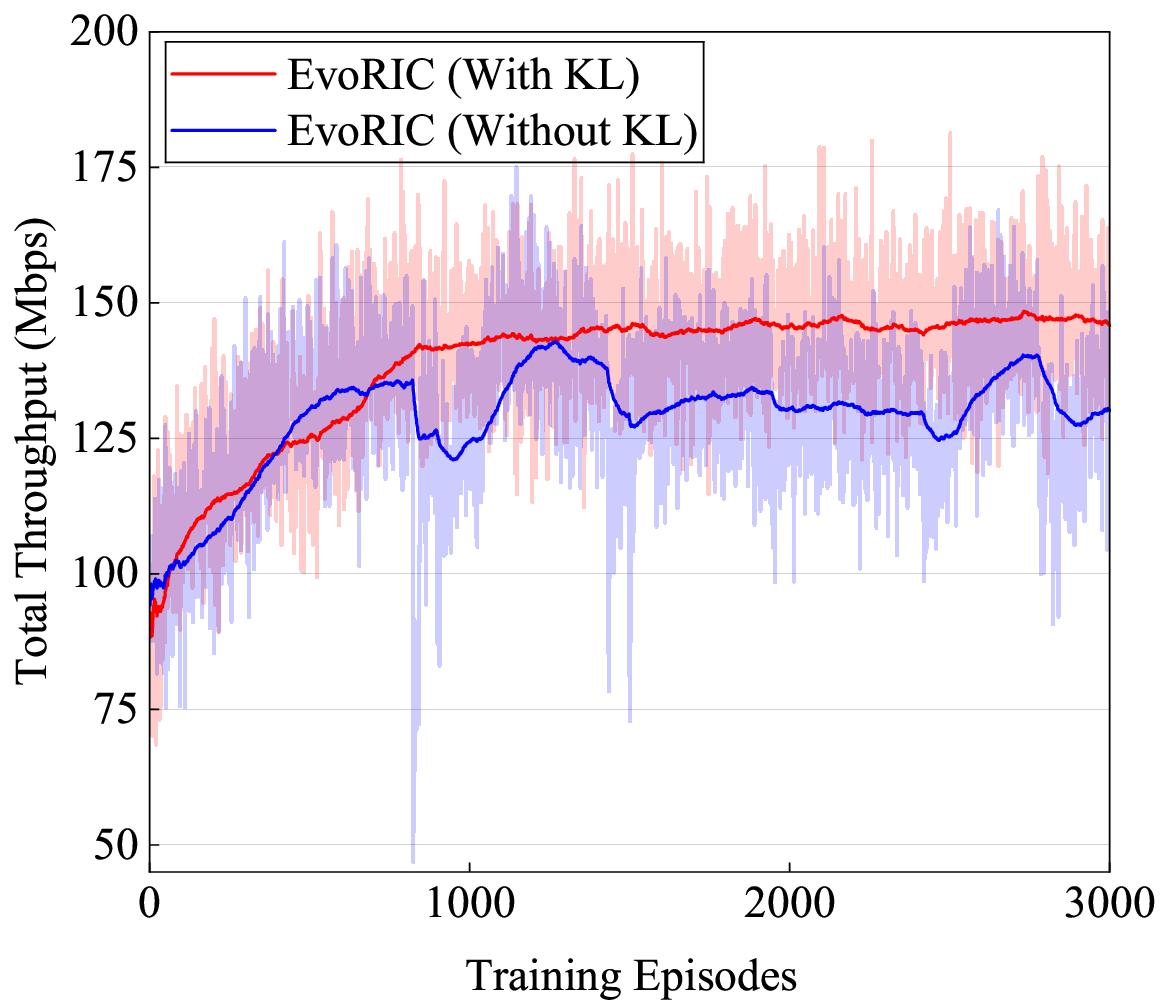}
	\caption{Training curve for EvoRIC with and without KL when $M=3, N=5, W=100\text{MHz}, P_m^{\text{max}}=44\text{dBm}$, and $\alpha=0.3$.}
	\label{fig:Traing_th}
\end{figure}

\blue{
Figure~\ref{fig:Traing_th} depicts the training dynamics of EvoRIC over 3,000 training episodes under a backhaul-access bandwidth partitioning ratio of $\alpha=0.3$. It compares the KL-regularized EvoRIC and EvoRIC trained without KL regularization. The KL-regularized configurations exhibit steady throughput improvement during the initial training stage and subsequently converge to stable performance. In contrast, EvoRIC without the KL penalty exhibits pronounced fluctuations, and excessive policy drift eventually causes numerical instability and interrupts further parameter updates. This comparison demonstrates the role of KL regularization in stabilizing PPO updates.
}

\begin{figure}[t]
	\centering
	\includegraphics[width=0.9\linewidth]{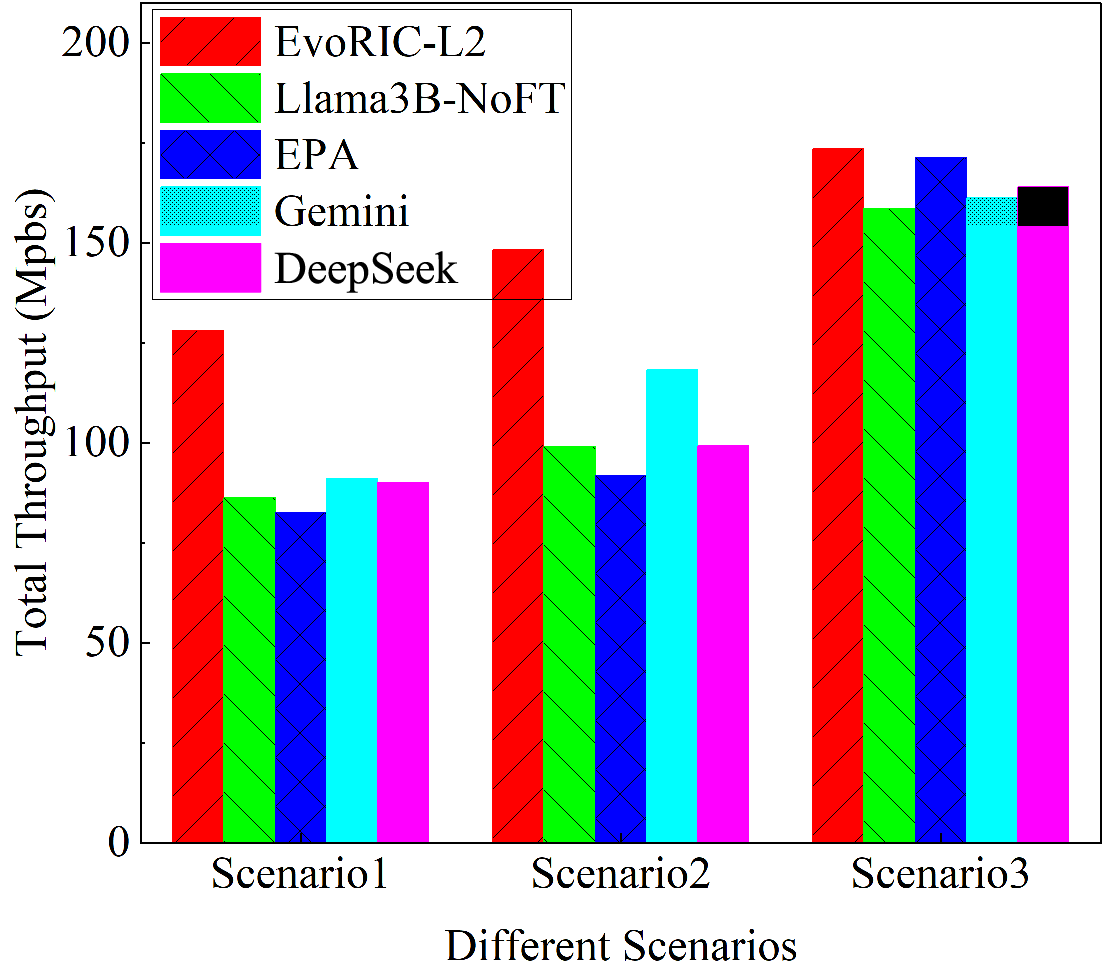}
	\caption{Average total throughput of the proposed EvoRIC and baseline methods according to different scenarios over 500 test examples.}
	\label{fig:scenario}
\end{figure}

\blue{Figure \ref{fig:scenario} illustrates the comparative throughput performance across three distinct network topologies: Scenario 1 ($M=3, N=6$), Scenario 2 ($M=3, N=5$), and Scenario 3 ($M=2, N=3$).
The results reveal that the proposed EvoRIC-L2 consistently outperforms all baselines, including the conventional \ac{EPA} scheme and the Llama3B-NoFT.
The performance gain is small in Scenario 3 because the optimization headroom is limited. With three SBSs per MBS, the solver improves upon EPA by only $16.6\,\%$, compared with $63$--$67\,\%$ in the other two scenarios, which means limited room for the learned policies to further outperform EPA.
This may occur when the end-to-end throughput is less affected by the backhaul power-allocation strategy or the optimal allocation is close to uniform.} 

\blue{From Fig.~\ref{fig:scenario}, we can also observe that EvoRIC achieves superior performance with only a fraction of the inference cost of general-purpose large models (i.e., DeepSeek and Gemini). The EvoRIC effectively adapts a compact, edge-deployable model to the specific characteristics of \ac{IAB} power allocation, while large models, despite possessing broad general knowledge, lack the specific `sense' for wireless resource constraints. Furthermore, EvoRIC completes inference in $0.56$ s, smaller than $3.82$ s of DeepSeek and $1.60$ s of Gemini, demonstrating its practical advantage for sub-second near-real-time control.}

\begin{table}[t]
\centering
\caption{Performance under non-nominal operating conditions in Scenario~2 over 500 paired episodes.} 
\label{tab:intent_fault}
\small
\setlength{\tabcolsep}{3pt}
\renewcommand{\arraystretch}{1.05}
\begin{tabular}{@{}lccc@{}}
\toprule
\shortstack[l]{Evaluation condition\\and reported metric}
& EvoRIC & Llama3B-NoFT & \shortstack{SCA-based \\algorithm} \\
\midrule
\multicolumn{4}{@{}l}{\textit{Intent adaptation}} \\
\shortstack[l]{PS (Mb/s)}
& 289.94 & 153.7 & 273.88 \\
\midrule
\multicolumn{4}{@{}l}{\shortstack[l]{\textit{Telemetry robustness:}}} \\
Fault-free (Mb/s)
& 166.57 & 81.69 & 155.45 \\
\shortstack[l]{URC} (Mb/s)
& 166.46 & 80.49	 & 150.58 \\
\bottomrule
\end{tabular}
\end{table}

\blue{Table~\ref{tab:intent_fault} presents a comparative analysis of intent adaptability and telemetry robustness in Scenario~2 over 500 paired episodes. For intent adaptability, the network topology, physical constraints, observed states, and action space remain unchanged. The same EvoRIC model fine-tuned for throughput maximization is evaluated without additional parameter updates, and only the intent description and evaluation metric are modified.}
\blue{For the priority-SBS (PS) intent, i.e., the throughput of the fourth SBS associated with each \ac{MBS} has a weight of five, while the remaining \acp{SBS} have unit weights. EvoRIC achieves the weighted throughput of $289.94$~Mb/s, outperforming EPA and the SCA-based algorithm by $96.6\%$ and $10.6\%$, respectively.}

\blue{For the telemetry robustness evaluation, the unmarked rate corruption (URC) is injected into the observations provided to the controllers, while the underlying physical environment remains unchanged. Specifically, the reported average rate of one \ac{SBS} per \ac{MBS} is multiplied by either $0.25$ or $4$ during the first two evaluated intervals.
No fault indicator is provided, and EvoRIC is evaluated without fault-aware parameter updates. Each prompt additionally contains the three preceding interaction records including the network state, executed action, and achieved throughput. As shown in Table~\ref{tab:intent_fault}, URC changes the throughput of EvoRIC slightly, from $166.57$ to $166.46$~Mb/s, whereas the throughput of the SCA-based algorithm decreases from $155.45$ to $150.58$~Mb/s, i.e., $3.1\%$ reduction. This result provides empirical evidence that EvoRIC can have better adaptability to different control objectives and be less sensitive to some information corruption, compared to the SCA-based algorithm. However, note that this does not establish general robustness to all telemetry faults.}

\section{Future Challenges} \label{sec:future challenges}
In this section, we discuss the open challenges for the EvoRIC framework in wireless communications. As illustrated in the previous sections, while \ac{RL} fine-tuning empowers \acp{LLM} for specific tasks, \blue{several challenges remain, including fine-grained real-time control, efficient exploration, multi-modal information processing, and automated lifecycle management.}

\subsection{Fine-grained and Real-time Control of Wireless Communication Resources}
A challenge for EvoRIC is bridging the granularity mismatch between discrete language tokens and continuous physical signals. \acp{LLM} operate on discrete tokens, creating precision barriers for continuous parameters like power levels. Current serialization schemes often result in precision loss or excessive context lengths. Future research may investigate semantic-aware tokenization or hybrid neuro-symbolic architectures, where the \ac{LLM} acts as a high-level planner while a lightweight controller handles fine-grained adjustments.

Moreover, edge deployment also introduces strict latency and resource constraints. While compact models (3B) are efficient for large-timescale management, symbol-level control demands millisecond responsiveness on limited hardware. \textit{Model distillation}, aggressive quantization (e.g., 4-bit), and novel architectures, such as large recurrent action models, may reduce the computational and memory overhead for deployment on COTS edge servers.

\subsection{Reasoning-Driven Exploration Strategies}
Currently, \ac{LLM} exploration relies primarily on the stochasticity of token generation (e.g., temperature sampling). However, this probability distribution is heavily biased by pre-trained linguistic knowledge, often leading to limited strategic diversity and convergence to suboptimal, `textually probable' solutions. \blue{Token-level randomness alone may therefore be insufficient for complex wireless state-action spaces.}

Therefore, future work should investigate reasoning-driven exploration strategies. This includes incorporating intrinsic motivation mechanisms, such as curiosity-driven exploration, to guide the \ac{LLM} toward novel policy regions. \blue{Furthermore, reflection-based methods (e.g., verbal reinforcement learning) may further enable the agent to analyze historical trajectories and low-reward actions, thereby combining semantic reasoning with conventional stochastic exploration.}

\subsection{Fine-Tuning with Multi-modal Input}
The current EvoRIC framework primarily processes serialized text-based inputs, which may discard the rich spatial and topological dependencies inherent in wireless environments (e.g., interference graphs, coverage maps). 
\blue{Consequently, text-only representations may limit the quality of resource-allocation decisions.}

A key future direction is extending the framework to support multi-modal fine-tuning, such as leveraging vision-language models. This involves developing robust encoding techniques to project diverse data types (vision, time-series, text) into a unified latent space, allowing the \ac{LLM} to reason over the full environmental context. By integrating multi-modal inputs, the \ac{RL} agent can better interpret complex spatial correlations and signal patterns that are difficult to describe via text alone, leading to more adaptive and informed control policies.

\subsection{Integration with Agentic AI for Automated Management}
Beyond optimizing the decision model, integrating \textit{agentic AI} to automate the RIC lifecycle represents a critical frontier. Manual tasks like reward engineering and dataset curation create significant bottlenecks requiring deep domain expertise.

Agentic AI frameworks can automate these processes by: 1) automated reward design: Synthesizing complex objective functions from high-level operator intents (e.g., `maximize fairness') and dynamically adjusting weights based on deployment feedback; and 2) autonomous data curation: Orchestrating digital twins to simulate diverse edge cases and automatically label high-quality trajectories for fine-tuning. 
\blue{Automating these processes may reduce the manual effort required to deploy and continuously evolve EvoRIC across heterogeneous \ac{O-RAN} environments.}

\section{Conclusions}
This paper proposed the EvoRIC framework, a novel architecture that integrates \acp{LLM} into \ac{O-RAN} for intelligent resource management. By structurally decoupling the system into a \ac{non-RT RIC} layer for \ac{RL}-based fine-tuning and a \ac{near-RT RIC} layer for \ac{near-RT} inference, the framework effectively addresses the challenges of adapting generative models to dynamic wireless environments. We evaluated EvoRIC for power allocation in multi-cell \ac{IAB} networks under different topologies and non-nominal operating conditions. Numerical results show that the compact fine-tuned model outperforms \ac{EPA}, Llama3B-NoFT, and the evaluated general-purpose models. The same model also adapts to changing control intents through prompt modification and exhibits limited degradation under transient telemetry corruption without additional parameter updates. These results demonstrate the potential of domain-specific fine-tuning for closed-loop and environment-driven \ac{O-RAN} control. Finally, we outlined critical future challenges in Section~\ref{sec:future challenges} to guide subsequent research in this emerging domain.

\bibliographystyle{IEEEtran}
\bibliography{Bib/IEEEabrv,Bib/ISCGroup,Bib/StringDefinitions,Bib/bib_LLM}

\end{document}